\documentclass[twocolumn,amsmath,amssymb,prc,superscriptaddress,floatfix,
showpacs]{revtex4}
\usepackage{graphicx}
\usepackage{color}
\usepackage{epsfig}
\usepackage[T1]{fontenc} 

\def\Eq#1{Eq.~(\ref{#1})}
\def\Fig#1{Fig.~\ref{#1}}

\def\Tab#1{Tab.~\ref{#1}}
\def\Sec#1{Sec.~\ref{#1}}
\newcommand{\vlowk}{\ensuremath{V_\text{low k}}}
\newcommand{\tlowk}{\ensuremath{T_\text{low k}}}

\newcommand{\eft}{\ensuremath{{\chi\text{EFT}\ }}}

\graphicspath{{./hyperfigures/}}

\begin{document}
\title{Hyperon-nucleon single-particle potentials with low-momentum
  interactions}

\author{H.~\DJ apo}
\email[E-Mail:]{haris@crunch.ikp.physik.tu-darmstadt.de}
\affiliation{Institut f\"ur Kernphysik, TU Darmstadt, 
Schlo{\ss}gartenstr.~9, D-64289 Darmstadt, Germany}
\author{B.-J. Schaefer}
\email[E-Mail:]{bernd-jochen.schaefer@uni-graz.at}
\affiliation{Institut f\"ur Physik, Karl-Franzens-Universit\"at Graz,
  Universit\"atsplatz 5, A-8010 Graz, Austria} 
\author{J. Wambach}
\affiliation{Institut f\"ur Kernphysik, TU Darmstadt, 
Schlo{\ss}gartenstr.~9, D-64289 Darmstadt, Germany}
\affiliation{Gesellschaft
  f\"ur Schwerionenforschung mbH, Planckstr.~1,  D-64291 Darmstadt, Germany}

\date{\today}

\begin{abstract}
  Single-particle potentials in Hartree-Fock approximation for
  different hyperon-nucleon ($YN$) channels are calculated in the
  framework of the effective low-momentum $YN$ interaction $\vlowk$.
  In contrast to the nucleon-nucleon interaction, the available
  experimental data for the $YN$ interaction are scarce. As a
  consequence no unique $YN$ low-momentum potential $\vlowk$ can be
  predicted from the various bare potentials. The resulting momentum-
  and density-dependent single-particle potentials for several
  different bare OBE models and for chiral effective field theory are
  compared to each other.
\end{abstract}

\maketitle


\section{Introduction}
\label{sec:intro}

The understanding of the hyperon-nucleon ($YN$) interaction is
essential to the physics of nuclear systems with strangeness and to
the octet of the lightest baryons. Furthermore, since e.g.~the core of
neutron stars may contain a high fraction of hyperons a deeper
knowledge of the $YN$ interaction is also important for astrophysical
issues. In recent years, an increased interest in exploring
nuclear systems with strangeness, especially multi-strange nuclear
systems, becomes apparent. The $YN$ interaction becomes relevant not
only when the properties of simple hypernuclei, double strange
hypernuclei but also when the production of hyperfragments in
relativistic heavy-ion collisions are studied. It also strongly
influences the composition and behavior of dense nuclear matter.
For an recent overview of hypernuclear physics see \cite{gal}.

Unfortunately, the details of the $YN$ interaction are known very
poorly. On the one hand, there exist only a very limited amount of
scattering data from which one could construct high-quality $YN$
potentials based on meson-exchange. The existing data do not constrain
the potentials sufficiently. On the other hand, theoretical analyses,
especially for many-body systems, do not seem to produce unambiguous
results \cite{kohno, vidana1}. For example, this uncertainty is
immediately demonstrated in that six different parameterizations of
the Nijmegen $YN$ potentials fit equally well the scattering data but
produce very different scattering lengths, see e.g.~\cite{rijken}. In
order to improve the reliability of available hyperon-nucleon and even
hyperon-hyperon potentials forthcoming experiments at the planned
J-PARC and FAIR facilities are indispensable. Recently, first lattice
QCD simulations on the $YN$ interaction have been performed
\cite{beane-2006}. In \cite{beane-2005-747} the findings on the
lattice concerning some aspects of $\Lambda N$ scattering are
confronted with results of the chiral effective field theory (\eft).

For the nucleon-nucleon interaction a unique low-momentum effective
potential is obtained \cite{Bogner:2003wn}. This uniqueness is lost in
the $YN$ case due to the less constraint bare potentials. Given that
there are significant differences between various $YN$ potentials it
will be interesting to see how those differences reflect themselves in
the hyperon single-particle potentials.

In this work we therefore wish to compare various $YN$ potentials and
determine their similarities and differences in a dense baryonic
medium. For this purpose we calculate the $YN$ single-particle
potentials and compare different low-energy scattering results. The
potentials are obtained in a Hartree-Fock approximation with an
effective low-momentum $YN$ potential $\vlowk$. The $\vlowk$ is
constructed from several bare $YN$ potentials \cite{schaefer, wagner}.
Since we work with isospin-symmetric nuclear matter, which contains no
hyperons we can neglect the hyperon-hyperon ($YY$) interaction. This
is advantageous since the $YY$ interaction is even less known and
constrained than the $YN$ interaction. It has been recently shown that
in the construction of the $YY$ interaction, at leading order \eft,
only one additional operator, which is not present in the $YN$
interaction, appears. The strength of this operator can be roughly
estimated from existing data on double hypernuclei
\cite{Polinder:2007mp}.

The outline of the paper is as follows: In the next Section we first
introduce the definition and formalism of the effective low-momentum
potential $\vlowk$ for the $YN$ interaction. The $\vlowk$ is obtained
as a solution of an renormalization group equation, which needs bare
$YN$ potentials as initial condition. We introduce and discuss in the
following some properties of the bare $YN$ potentials, used in this
work and present some low-energy scattering quantities. The following
section, \Sec{sec:Single-particle potentials}, is devoted to the
calculation of the single-particle potentials in the Hartree-Fock
approximation. Results for symmetric nuclear matter, the partial-wave
contribution to the single-particle potentials and a comparison to
other approaches are shown. Finally, we end in \Sec{sec:summary} with
a summary and conclusion.

\section{Low-momentum interactions}
\label{sec:vlowk}

We consider hyperons ($Y = \Lambda, \Sigma^+, \Sigma^0, \Sigma^-$)
with strangeness $S=-1$ in an infinite system of nucleons $N$ composed
of equal numbers of protons $p$ and neutrons $n$. We consider
densities around nuclear saturation density $\rho_0 \approx 0.16$
fm$^{-3}$ corresponding to a Fermi momentum $k_{F,0}= 1.35$ fm$^{-1}$.
The electromagnetic interaction is switched off. There exists only a
very limited amount of scattering data with which one can construct
phenomenological potentials. These potentials are used as initial
condition for solving RG equations, which lead to effective
low-momentum $YN$ interactions $\vlowk$'s. From these different
effective $\vlowk$'s we finally calculate the single-particle
potential for the $\Lambda$-hyperon. We start with a brief review of
the construction of the low-momentum $YN$ interaction. Details for
several bare $YN$ potentials within the RG framework used here can be
found in Refs.~\cite{schaefer, wagner}.

\subsection{Construction of \vlowk}
\label{sec:formalism}

The starting point for the construction of the $\vlowk$ is the
half-on-shell $T$-matrix, $T(q',q;E_y )$, which is determined by the
nonrelativistic Lippmann-Schwinger equation in momentum space. The
on-shell energy is denoted by $E_y$ while $q'$, $q$ are relative
momenta between a hyperon and nucleon. An effective low-momentum
$\tlowk$-matrix is then obtained by introducing a cutoff $\Lambda$ in
the Lippmann-Schwinger kernel thus integrating the intermediate-state
momenta up to this cutoff. At the same time, the bare potential in the
coupled-channel partial-wave Lippmann-Schwinger equation are replaced
with the corresponding low-momentum potential $\vlowk$,
\begin{eqnarray}
  & &T_{\text{low }k, y' y}^{\alpha'\alpha}(q',q; E_y )=
    V_{\text{low }k,  y' y}^{\alpha'\alpha}(q',q)+ \nonumber \\
  &&\frac{2}{\pi}
  \sum_{\beta, z} P\!\!\int\limits_0^{\Lambda}\!\!dl\; l^2
  \frac{V_{\text{low }k, y' z}^{\alpha'\beta}(q',l)
        T_{\text{low }k, z
          y}^{\beta\alpha}(l,q;E_y)}{E_{y}(q)-E_{z}(l)}\ .
\label{eq:vlowk} 
\end{eqnarray}
The labels $y$, $z$ indicate the particle channels, e.g.~$y=YN$ and
$\alpha$, $\beta$ denote the partial waves, e.g.~$\alpha=LSJ$ where
$L$ is the angular, $J$ the total momentum and $S$ the spin. In
Eq.~(\ref{eq:vlowk}) the energies are given by
\begin{eqnarray}
E_{y}(q)&=&M_{y}+\frac{q^2}{2\mu_{y}},
\label{eq:ene} 
\end{eqnarray}
with the reduced mass $\mu_{y}=M_Y M_N/M_y$ and total mass
$M_{y}=M_Y+M_N$ of the hyperon $M_Y$ and the nucleon $M_N$.

Finally, the effective low-momentum $\vlowk$ is defined by the
requirement that the $T$-matrices are equivalent for all momenta below
this cutoff
\begin{equation*} 
T^{\alpha'\alpha}(q',q; E) = \tlowk^{\alpha'\alpha}(q',q; E)\ , 
\qquad q',q \leq \Lambda\ .
\end{equation*}
The $\vlowk$ thus obtained is non-hermitian but nevertheless phase-shift
equivalent hermitian low-momentum $YN$ interactions can be obtained.
Since the low-momentum $T$-matrix, $\tlowk$ must be
cutoff-independent, i.e. $d\tlowk/d\Lambda = 0$ an RG flow equation
for $\vlowk$
\begin{equation}
  \frac{d\vlowk(k',k)}{d\Lambda} = \frac{2}{\pi}\frac{\vlowk(k',\Lambda)
       T(\Lambda,k;\Lambda^2)}{1-k^2/\Lambda^2}
  \label{eq:vlowk_flow}
\end{equation}
can immediately be derived. Instead of solving this flow equation with
standard numerical methods (e.g.~Runge-Kutta method) directly, the
so-called ALS iteration method, pioneered by Andreozzi, Lee and
Suzuki, is used~\cite{lee80,suzuki80,andreozzi}. This iteration method
is based on a similarity transformation and its solution corresponds
to solving the flow equation. Details about the convergence of the ALS
iteration method, applied for the coupled channel $YN$ interaction,
can be found in \cite{wagner}. For the hyperon-nucleon interaction
with strangeness $S=-1$ two different basis systems, the isospin and the
particle basis of the bare potentials are available. We will use the
latter.

Furthermore, in our investigations only the diagonal elements of the
matrices are needed. Therefore, we will shorten our notation further
and introduce the abbreviation
$V^{\alpha'\alpha}_{y'y}(q',q)\rightarrow V^{\alpha}_{y}(q)$ for all
diagonal quantities.

\subsection{Bare potentials}
\label{sec:propertiesYN}

In order to solve the flow equation (\ref{eq:vlowk_flow}) a bare
potential as initial condition for the flow has to be chosen. In this
work several initial $YN$ potentials, the original Nijmegen soft core
model NSC89 \cite{nsc89}, the series of models NSC97a-f \cite{rijken}
also by the Nijmegen group and a recent model proposed by the J\"ulich
group \cite{juelich}, labeled as J04 in the following, are used. All
these models are formulated in the conventional meson-exchange (OBE)
framework. They involve a set of parameters, which have to be
determined from the available scattering data. These are the coupling
constants of the corresponding baryon-baryon-meson vertices and cutoff
parameters for the vertex form factors. Due to the scarce $YN$
scattering data base these parameters cannot be precisely fixed in
contrast to the $NN$ interaction, where a lot of scattering data are
available. In order to construct consistently conventional OBE models
for the $YN$ interaction one usually assumes flavor $SU(3)$
constraints or $G$-parity arguments on the coupling constants, and in
some cases even the $SU(6)$ symmetry of the quark model and adjusts
their size by fits to $NN$ data. The major conceptual difference
between the various conventional OBE models consists in the treatment
of the scalar-meson sector, which plays an important role in any
baryon-baryon interaction at intermediate ranges. In contrast to the
pseudoscalar and vector meson sectors it is still an open issue who
are the actual members of the lowest-lying scalar-meson $SU(3)$
multiplet, what are the masses of the exchange particles and how, if
at all, the relations for the coupling constants, obtained by $SU(3)$
flavor symmetry, should be applied. For example, in the older versions
of the $YN$ models by the J\"ulich group \cite{Holzenkamp:1989tq,
  Reuber:1993ip} a fictitious $\sigma$ meson with a mass of roughly
550 MeV, arising from correlated $\pi\pi$ exchange was introduced. The
coupling strength of this meson to the baryons was treated as a free
parameter and finally fitted to the data. However, in the novel
J\"ulich $YN$ potential \cite{juelich} a microscopic model of the
correlated $\pi\pi$ and $K \bar K$ exchange is established in order to
fix the contributions in the scalar $\sigma$- and vector
$\rho$-channel. This new model incorporates also the common one-boson
exchange parts of the lowest pseudoscalar and vector meson multiplets.
The corresponding coupling constants are determined by $SU(3)$ flavor
symmetry and the so-called $F/(F+D)$ ratios are fixed by the
pseudoscalar and vector meson multiplets by invoking $SU(6)$ symmetry.

In the Nijmegen $YN$ models, NSC89 \cite{nsc89}, NSC97 \cite{rijken}
and in the recently extended soft-core model for strangeness $S=-2$
ESC04 \cite{Rijken:2006kg, Rijken:2006ep} the interaction is
generated by a genuine scalar $SU(3)$ nonet meson exchange. Besides
the scalar-meson nonet two additional nonets, the pseudoscalar and 
vector $SU(3)$ flavor nonets are considered in all Nijmegen models.
In addition, Pomeron exchange is also included, which provides 
additional short-range repulsion. Nevertheless, there are a few
conceptual differences in the various Nijmegen models. In the NSC97
models the strength parameter for the spin-spin interaction the
magnetic $F/(F+D)$ ratio is left as an open parameter and takes six
different values in between a range of $0.4447$ to $0.3647$ for the
six different models NSC97a-f. In the original Nijmegen SC89 model
this parameter is constrained by weak-decay data. Furthermore, the
NSC97 models include additional $SU(3)$ flavor breaking, which is based
on the so-called $^3$P$_0$ model \cite{micu}.

The predictions of the above mentioned models can be compared with an
another approach, the so-called chiral effective field theory (\eft)
to nuclear interactions, which is based on chiral perturbation theory
(for recent reviews see e.g.~\cite{bedaque-2002-52, epelbaum-2006-57,
  Furnstahl:2008df}). The major benefit of the \eft is the underlying
power counting scheme, proposed by Weinberg \cite{Weinberg:1990rz,
  Weinberg:1991um}, that allows to improve the calculations
systematically by going to higher orders in the expansion. In
addition, higher two- and three-body forces can be derived
consistently in this framework. Furthermore, the effective potential
is explicitly energy independent in contrast to the original Weinberg
scheme.

Within \eft the $NN$ interaction has been analyzed recently to a high
precision (N$^3$LO) \cite{epelbaum-2005-747}. To leading order (LO)
the $NN$ potential is composed of pion exchanges and a series of
contact interactions with an increasing number of derivatives which
parameterize the singular short-range part of the $NN$ force. In order
to remove the high-energy components of the baryonic and pseudoscalar
meson fields a cutoff $\Lambda$ dependent regulator function in the
Lippmann-Schwinger (LS) equation is introduced. With this
regularized LS equation observable quantities can be calculated. The
cutoff range is limited from below by the mass of the pseudoscalar
exchange mesons. Note, that in conventional meson-exchange models the
LS equation is not regularized and convergence is achieved by
introducing form factors with corresponding cutoff masses for each
meson-baryon-baryon vertices.

So far, the $YN$ interaction has not been investigated in the context
of the \eft as extensively as the $NN$ interaction. A recent
application to the $YN$ interaction by the J\"ulich group can be found
e.g.~in \cite{poli}. Analogous to the $NN$ case, the $YN$ potential,
obtained in LO \eft, and consists of four-baryon contact terms and
pseudoscalar meson (Goldstone boson) exchanges, which are all related
by $SU(3)_f$ symmetry. For the $YN$ interaction typical values for the
cutoff lie in the range between 550 and 700 MeV (see
e.g.~\cite{epelbaum-2005-747}). At LO \eft and for a fixed cutoff
$\Lambda$ and pseudoscalar $F/(F+D)$ ratio there are five free
parameters. The remaining interaction in the other $YN$ channels are
then determined by $SU(3)_f$ symmetry. A next-to-leading order (NLO)
\eft analysis of the $YN$ scattering and of the hyperon mass shifts in
nuclear matter was performed in \cite{korpa-2002-65}. However, in this
analysis the pseudoscalar meson-exchange contributions were not taken
into account explicitly but the $YN$ scattering data could be
described successfully for laboratory momenta below 200 MeV using 12
free parameters. One ambiguity in this approach is the value of the
$\eta$ coupling which is identified with the octet $\eta_8$ meson
coupling and not with the physical $\eta$ meson. The influence of this
ambiguity on the data description can be disregarded
\cite{Haidenbauer:2007ra}.

Since there are scarce $YN$ scattering data available, it has not been
possible yet to determine uniquely the spin structure of the $YN$
interaction. Nevertheless, all of the above mentioned OBE models are
consistent with the measured $YN$ scattering observables.
In addition, all of these potentials include the
$\Lambda N-\Sigma N$ conversion process.

\subsection{Low-energy scattering}
\label{sec:scattering length}

In order to obtain further insight into the separation of scales for
the evolution of the low-momentum $\vlowk$ we investigate its cutoff
dependence. A common feature of all $YN$ potentials is the long-range
one-pion exchange (OPE) tail. In general, the RG decimation eliminates
the short-distance part of the bare potential and preserves the
model-independent impact of the high-momentum components on
low-momentum observables. In this sense, the ambiguities associated
with the unresolved short-distance parts of the interaction disappear
and a universal low-momentum $YN$ interaction $\vlowk$ can be
constructed from phase shift equivalent bare $YN$ potentials.

\begin{figure}[!htb]
  \centerline{\hbox{
      \includegraphics[width=\linewidth]{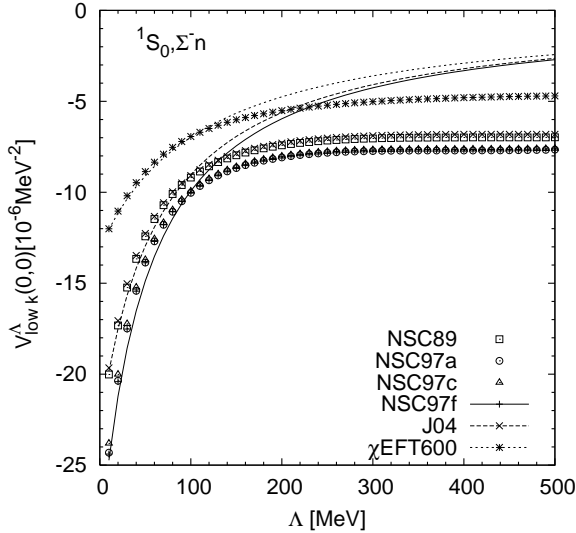}
    }}
  \caption{\label{fig:running} $\vlowk^\Lambda (0)$ in $^1S_0$ partial
    wave for various bare potentials as a function of the cutoff
    $\Lambda$ in the $\Sigma^- n$ channel. Prediction from effective
    range theory (lines) are added.}
\end{figure}

The mentioned hierarchy of scales can be seen e.g.~in the $\Sigma^- n$
channel, see \Fig{fig:running}. The $\vlowk$ matrix elements for
vanishing momenta are shown as a function of the cutoff $\Lambda$ for
the $^1S_0$ partial wave. When $\Lambda$ is decreased, the resulting
$\vlowk$ becomes more and more attractive. For $^1S_0$ and a cutoff
$\Lambda \sim 500 - 250$ MeV $\vlowk$ becomes cutoff independent.
Decreasing the cutoff further below the $2\pi$ exchange threshold,
which corresponds to a $k \approx 280$ MeV, the cutoff insensitivity
disappears since the pion contributions are finally integrated out.

In the opposite direction, i.e.~for $\Lambda \to \infty$ no
fluctuations have been integrated and $\vlowk$ tends to the bare
potential.

The limit $\Lambda \to 0$ of $\vlowk$ should yield the scattering
length. In the limit of small cutoffs an analytic solution obtained in
the framework of the effective theory, see \cite{kaplan}, is given by
the expression
\begin{eqnarray}
  \label{eq:flow}   
  V_{y}(0)=\left[2\frac{\mu_{y}}{a_0}-2\frac{\Lambda}{\pi}\right]^{-1}
  \qquad \mbox{for} \qquad \Lambda \rightarrow 0\ ,
\end{eqnarray} 
where we have simplified our notation in an obvious manner. Here, the
scattering length $a_0$ is needed as input, which we have calculated in
the standard effective range approximation directly from the
$T$-matrix for the $^1S_0$ channel from the $\vlowk$. In this
approximation the $T$-matrix for $q \leq \Lambda$ can be expanded as
\begin{eqnarray}
  \label{eq:scat} 
  q \cot \delta_0 = -\frac{1}{2\mu_{y} T_{y}(q,q;q^2)}=-\frac{1}
  {a_0}+\frac{1}{2}r_0 q^2\ ,
\end{eqnarray}
where $r_0$ is the effective range. The results for all different $YN$
flavor channels and for all bare OBE potentials and the \eft potential
used in this work with cutoffs between $550$ and $700$ MeV are listed
in \Tab{tab:Ssl} for the scattering length $a_0$ in units of fm
\begin{table}[h!]
\begin{center}
\begin{tabular}{|c||c|c|c|c|c|c|c|c|}
\hline
       &$\Lambda p$&$\Lambda n$&$\Sigma^0 p$&$\Sigma^0 n$
       &$\Sigma^+ p$&$\Sigma^+ n$&$\Sigma^- p$&$\Sigma^- n$ \\ \hline\hline
NSC97a & -0.71 & -0.76 & -2.46 & -1.74 & -6.06 & -0.04 & 0.41 & -6.13 \\ \hline
NSC97b & -0.90 & -0.96 & -2.47 & -1.72 & -5.98 & -0.04 & 0.41 & -6.06 \\ \hline
NSC97c & -1.20 & -1.28 & -2.41 & -1.70 & -5.90 & -0.03 & 0.41 & -5.98 \\ \hline
NSC97d & -1.70 & -1.82 & -2.38 & -1.68 & -5.82 & -0.03 & 0.41 & -5.89 \\ \hline
NSC97e & -2.10 & -2.24 & -2.38 & -1.68 & -5.82 & -0.03 & 0.41 & -5.90 \\ \hline
NSC97f & -2.51 & -2.68 & -2.45 & -1.74 & -6.07 & -0.05 & 0.42 & -6.16 \\ \hline
NSC89  & -2.70 & -2.72 & -2.12 & -1.57 & -4.79 & -0.09 & 0.23 & -4.85 \\ \hline
J04    & -2.14 & -2.11 & -2.24 & -1.63 & -4.68 & -0.18 & 0.04 & -4.75 \\ \hline
$\chi$EFT550& -1.80 & -1.79 & -1.76 & -1.15 & -3.82 & 0.12 & 0.31 & -3.88 \\ \hline
$\chi$EFT600& -1.80 & -1.80 & -1.25 & -0.92 & -2.70 & 0.10 & 0.20 & -2.72 \\ \hline
$\chi$EFT650& -1.80 & -1.80 & -1.43 & -1.02 & -3.06 & 0.09 & 0.21 & -3.10 \\ \hline
$\chi$EFT700& -1.80 & -1.80 & -1.50 & -1.07 & -3.19 & 0.06 & 0.20 & -3.24 \\ \hline
\end{tabular}                               
\caption{Scattering lengths $a_0$ of $\vlowk$ for different flavor
  channels in units of $\text{fm}$ for the $^1S_0$ partial wave.}
  \label{tab:Ssl}
\end{center}
\end{table}
and in \Tab{tab:Ser} for the effective range $r_0$ also in units of
fm.

\begin{table}[h!]
\begin{center}
\begin{tabular}{|c||c|c|c|c|c|c|c|c|}
\hline
       &$\Lambda p$&$\Lambda n$&$\Sigma^0 p$&$\Sigma^0 n$
       &$\Sigma^+ p$&$\Sigma^+ n$&$\Sigma^- p$&$\Sigma^- n$ \\ \hline\hline
NSC97a & 5.87 & 6.12 & 4.58 & 0.60 & 3.28 & -6602 & 24.8 & 3.27      \\ \hline
NSC97b & 4.93 & 5.10 & 4.68 & 0.59 & 3.29 & -8491 & 25.0 & 3.28      \\ \hline
NSC97c & 4.11 & 4.23 & 4.79 & 0.57 & 3.30 & -10670 & 25.4 & 3.29      \\ \hline
NSC97d & 3.46 & 3.53 & 4.91 & 0.54 & 3.30 & -17115 & 25.4 & 3.29      \\ \hline
NSC97e & 3.19 & 3.24 & 4.90 & 0.52 & 3.29 & -17326 & 25.2 & 3.29      \\ \hline
NSC97f & 3.03 & 3.09 & 4.60 & 0.51 & 3.25 & -6341 & 24.1 & 3.24      \\ \hline
NSC89  & 2.86 & 2.98 & 5.76 & 0.74 & 3.35 & -1478 & 58.0 & 3.33      \\ \hline
J04    & 2.93 & 3.09 & 3.76 & 1.04 & 3.32 & -329 & 1232.0 & 3.30      \\ \hline
$\chi$EFT550& 1.73 & 1.84 & 6.10 & -2.96 & 2.70 & -825 & 34.1 & 2.68 \\ \hline
$\chi$EFT600& 1.77 & 1.88 & 5.32 & -2.12 & 3.40 & -780 & 10.2 & 3.39 \\ \hline
$\chi$EFT650& 1.75 & 1.86 & 5.10 & -2.28 & 3.08 & -1210 & 27.6 & 3.05 \\ \hline
$\chi$EFT700& 1.74 & 1.86 & 4.91 & -2.17 & 2.97 & -2450 & 34.8 & 2.95 \\ \hline
\end{tabular}                               
\caption{Effective range $r_0$. Labeling is the same as in \Tab{tab:Ssl}.}
  \label{tab:Ser}
\end{center}
\end{table}

As can be seen from \Fig{fig:running} there is good agreement for small
cutoff values $\Lambda$ between the analytical expansion and the full
$\vlowk$ solution obtained from the flow equation.

Unfortunately, no general quantitative conclusion can be drawn from
\Tab{tab:Ssl} and \Tab{tab:Ser} due to the bad experimental situation
for the $YN$ data. The $YN$ interaction is yet largely unknown.
However, agreement of the scattering length of all NSC97
potentials except for the $\Lambda p$ and $\Lambda n$ channels is
found. This deviation is related to the different fits of the magnetic
$F/(F+D)$ ratio in the Nijmegen potentials \cite{rijken}. The
remaining two potentials, NSC89 and J04, have different but comparable
values to those of the NSC97 ones. Unfortunately, the difference
between these potentials and the \eft potential is large.

\begin{figure}[!htb]
  \centerline{\hbox{
      \includegraphics[width=\linewidth]{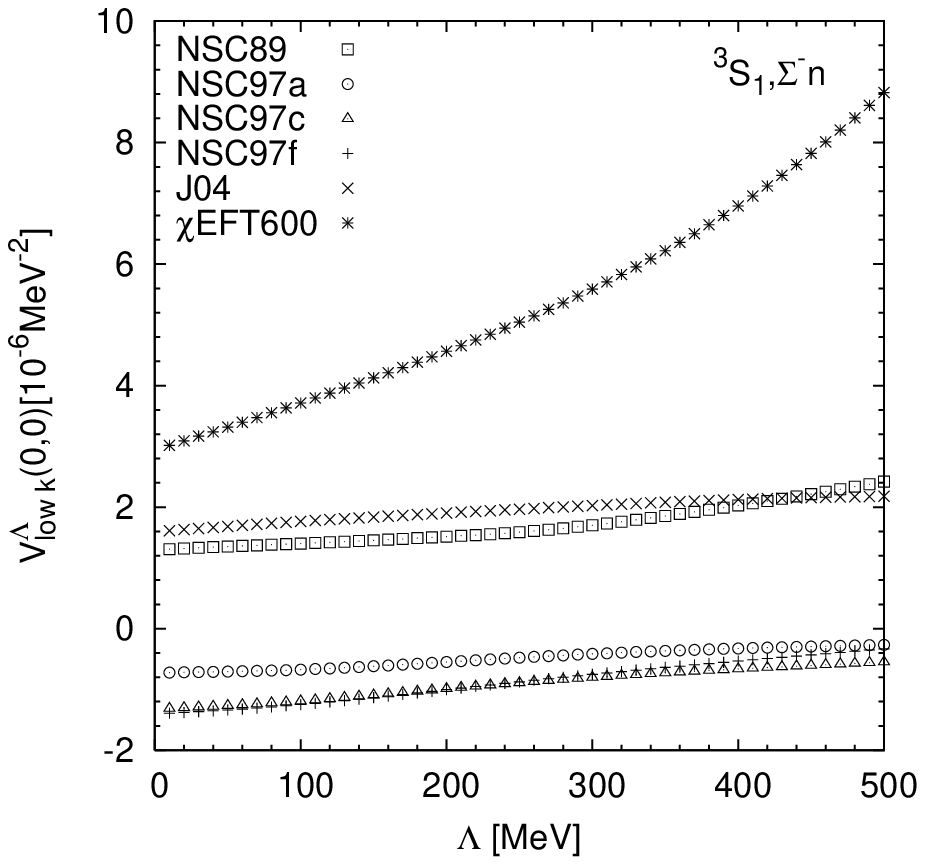}
    }}
  \caption{\label{fig:running3} $\vlowk^\Lambda (0)$ similar to
    \Fig{fig:running} for the $^3S_1$ channel.}
\end{figure}

\Fig{fig:running3} shows the same matrix element as in
\Fig{fig:running} but for the $^3S_1$ partial wave. Unlike to the
$^1S_0$ channel, $\vlowk$ for the $^3S_1$ channel is still cutoff
dependent. This can be traced back to the different short-range
behavior of the two channels. In the $^1S_0$ channel the potential has
an strongly repulsive core while in the $^3S_1$ channel the
short-distance part is strongly attractive. Hence, during the RG
decimation towards smaller cutoff values, the potential gets more and
more attractive.

\section{Single-particle potentials}
\label{sec:Single-particle potentials}

Generally, the single-particle potential $U(p)$ is defined as the
diagonal part in spin and isospin space of the proper self-energy for
the single-particle Green's function. In the Hartree-Fock
approximation for a uniform system it represents the first-order the
interaction energy of a particle with incoming momentum $p$ and given
spin and isospin with the filled Fermi sea. For the $YN$ interaction
the hyperon-nucleon single-particle potential $U_Y(p)$ describes the
behavior of the incoming hyperon $Y$ with momentum $p$ in the nuclear
medium, i.e.~its interaction with a filled Fermi sea of nucleons. The
ground-state energy of nuclear matter in first approximation is then
obtained by integrating $U(p)$ over the incoming momentum, spin and
isospin and adding the kinetic energy. Pictorially, the ground-state
energy is represented by closed Goldstone diagrams. By cutting one
line, symbolizing the hyperon propagator, in each corresponding
Goldstone diagram the $U_Y(p)$ is obtained.

In the following, the calculation of the momentum- and
density-dependent $U_Y(p)$ in the Hartree-Fock approximation is
presented for symmetric nuclear matter.

\subsection{Hartree-Fock approximation}
\label{sec:results}

The single-particle potential $U_Y(p)$ for a hyperon with momentum
$p = |\vec p |$ is obtained from the diagonal elements of the
low-momentum potential matrix, $V^\alpha_{y}(q)$, where the labeling
introduced in \Eq{eq:vlowk} has been used. In a plane-wave basis it
splits into two terms, the (direct) Hartree- and the (exchange)
Fock-term~\cite{Walecka}.
\begin{eqnarray}
V^\alpha_{y}(q)=\left.V^{\alpha}_{y}(q)\right|_{\rm direct}+
(-1)^{L+S} \left.V^{\alpha}_{y}(q)\right|_{\rm exchange}~.
\end{eqnarray}
Since the low-momentum potential is given in
the partial-wave basis we perform a change of basis with the result

\begin{equation}
  U_{Y}(p)=\!\!\!\!\!\sum_{\alpha,M_S,M_L}\!\!\!f^{\alpha}
  \sum_{N} \int\limits_{t_{min}}^{t_{max}}\!\!\!dt P^2_{LM_L}(t)\!\!\!
  \int\limits_{q_{min}}^{q_{max}}\!\!\! dq q^2 V^{\alpha}_{y}(q)\ ,
\label{eq:spp}
\end{equation}
where $\vec{q} = ({M_Y\vec{k}-M_N\vec{p}})/M_{y}$ denotes the relative
momentum between the nucleon $\vec k$ and hyperon momentum $\vec p$.
The $P_{LM_L}(t)$ are
the associated Legendre polynomials of the first kind with the
argument $t=\hat{\vec{p}}\cdot\hat{\vec{q}}$, where the hat labels an
unit vector. The quantity $f^{\alpha}$ is the short-hand notation for
the square of the Clebsch-Gordon coefficients
\begin{eqnarray}
f^{\alpha}=\left(\begin{array}{cc|c}L & S & J\\ M_L& M_S&
M_J\end{array}\right)^2
\frac{2L+1}{2\pi}\frac{(L-M_L)!}{(L+M_L)!}\ ,
\label{eq:pre}
\end{eqnarray}
and the $M_J$, $M_L$ and $M_S$ are the corresponding projections of
the total $J$, angular momentum $L$ and spin $S$, respectively.

The integration boundaries of $U_Y(p)$ in \Eq{eq:spp} are derived in
the following: For vanishing temperature the momentum of the nucleon
$\vec k$ is restricted by the Fermi momentum $\vec k_F$, i.e.
$0\leq k \leq k_F$, and $k_F$ is directly related to the proton or
neutron density $\rho_N$ via
\begin{equation}
\label{eq:fermimom} 
k_{F}^3= 3\pi^2x_N\rho_B\ .
\end{equation}
Here $x_N=\rho_N /{\rho_B}$ describes the ratio between the proton or
neutron and total baryon density. The inequality $0\leq k\leq k_F$ can be
reformulated as
\begin{equation*}
M^2_{y}q^2+M^2_{N}p^2+2M_{y}M_{N}qpt-M^2_Yk^2_F\leq 0 
\label{eq:qinequ}
\end{equation*}
which has the solution for the relative momentum $q$
\begin{equation*}
q^- (k_{F},p,t) \leq q \leq q^+ (k_{F},p,t)
\end{equation*}
with the definitions
\begin{equation}
\label{eq:border3}
q^{\pm}(k_{F},p,t) = \frac{M_N}{M_{y}}\!\left[ p\cdot t \pm
  \sqrt{ \frac{M_Y^2}{M_N^2}k^2_{F}-p^2(1-t^2)}\right]\ .
\end{equation}

Since the relative momentum $q$ is a real quantity this further
constrains the integration variable $t$ to
\begin{equation}
t\geq \sqrt{1-\left(\frac{M_Y}{M_N}\frac{k_{F}}{p}\right)^2}
\label{eq:tlimit} 
\end{equation}  
which is only valid if the hyperon momentum is
$p\geq \frac{ M_Y}{M_N}k_F$. In this case this finally determines the
integration limits as
\begin{align}
  t_{min}=&\sqrt{1-\left(\frac{M_Y}{M_N}\frac{k_{F}}{p}\right)^2}&;\quad
  &t_{max}=1\ ;\nonumber \\
  q_{min}=&q^-(k_{F},p,t)&;\quad &q_{max}=q^+(k_{F},p,t)
\label{eq:border2}
\end{align}
because the modulus of $t$ is always smaller or equal one.

For the case that the hyperon momenta $p\leq \frac{ M_Y}{M_N}k_F$ the
functions $q^\pm (k_F, p, t)$ are always real which then yield the
integration limits
\begin{align}
t_{min}=&-1&;\quad &t_{max}=1\ ;\nonumber \\
q_{min}=&0 &;\quad &q_{max}=q^+(k_{F},p,t)\ .
\label{eq:border1}
\end{align}

\begin{figure}[!htb]
  \centerline{\hbox{
      \includegraphics[width=\linewidth]{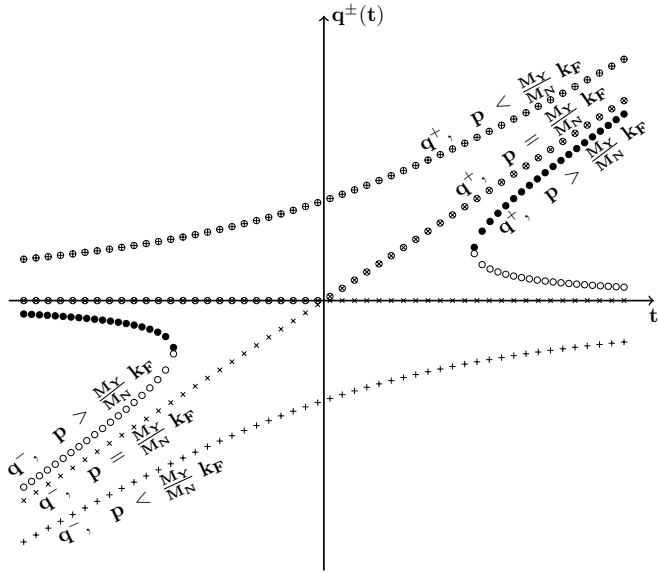}
    }}
  \caption{\label{fig:border1} $q^{\pm}(k_{F},p,t)$ as function of $t$
    for different choices of $p$ and fixed $k_{F}$.}
\end{figure}

In Fig.~\ref{fig:border1} the integration limit functions $q^\pm$ are
shown as a function of $t$ for three different choices of the hyperon
momentum
($p< \frac{M_Y}{M_N}k_F,p> \frac{M_Y}{M_N}k_F, p= \frac{M_Y}{M_N}k_F$)
and a fixed $k_F$.

Thus, the integration limits are known analytically and finally the
$U_Y(p)$ are calculated via \Eq{eq:spp} numerically with standard
integration methods.

\begin{figure}[!ht]
  \centerline{\hbox{
      \includegraphics[width=1.5\linewidth]{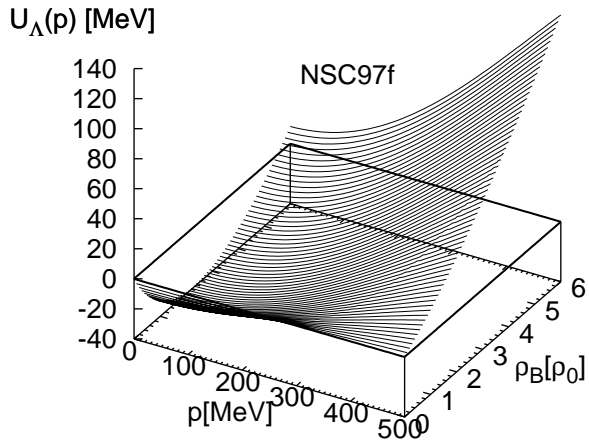}}}
  \caption{Momentum and density dependence of $U_\Lambda (p)$ for
    symmetric nuclear matter. As bare potential the NSC97f was used.
  }
  \label{fig:LkNf}
\end{figure}

\begin{figure}[!ht]
  \centerline{\hbox{
      \includegraphics[width=1.5\linewidth]{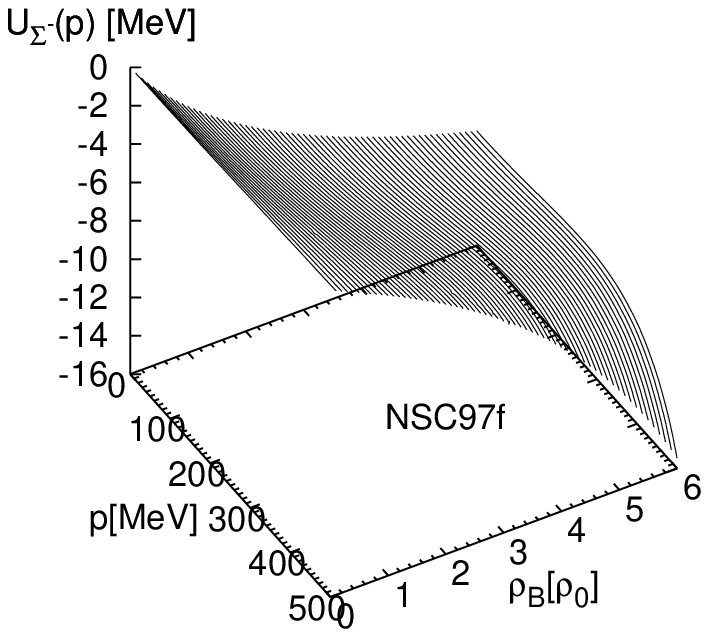}}}
  \caption{Similar to \Fig{fig:LkNf} for $U_{\Sigma^-} (p)$.}
  \label{fig:SmNf}
\end{figure}

\subsection{Symmetric nuclear matter}
\label{sec:Symmetric nuclear matter results}

For symmetric nuclear matter the ratio in \Eq{eq:fermimom} has to be
fixed to $x_N = 1/2$. As an example, the numerical solution of
\Eq{eq:spp} for the full momentum and density dependent
single-particle potential of the $\Lambda$ hyperon with momenta up to
500 MeV and nuclear densities up to 6$\rho_0$ is shown in
\Fig{fig:LkNf} where as bare potential for the underlying $\vlowk$
calculation the NSC97f model of the Nijmegen group has been used. One
sees that with increasing density, the momentum dependence becomes
stronger, indicating that the effective mass becomes lower for larger
densities.

Similarly, \Fig{fig:SmNf} shows the full momentum and density
dependence of the $\Sigma^-$ potential in symmetric nuclear matter
based on the NSC97f potential. Here, the slope of the momentum
dependence is less pronounced, which leads to a weaker density
dependent effective mass. However, unlike to the $\Lambda$ case, the
curvature becomes negative at higher densities leading to an larger
effective mass than the bare mass.

\begin{figure}[!htb]
  \centerline{\hbox{
      \includegraphics[width=1.3\linewidth]{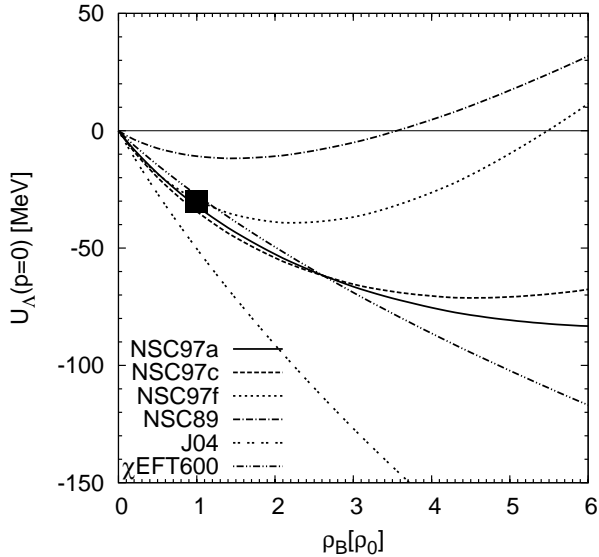}}}
  \caption{\label{fig:Lrho} $U_{\Lambda} (p=0)$ as a function of density
    in symmetric nuclear matter. The square represents the
    empirical point $U_{\Lambda}(p=0)\approx -30\; \text{MeV}$
    \cite{saha}.}
\end{figure}

The density dependence for several $\Lambda$ potentials at rest (i.e.
$p=0$) in symmetric nuclear matter is compared in \Fig{fig:Lrho}. The
square represents the generally excepted empirical potential depth of
$U_{\Lambda}(p=0)\approx -30\;\text{MeV}$. This value has been
recently confirmed by an analysis of the ($\pi^-,K^+$) inclusive
spectra on various target nuclei as best fits in a framework of a
distorted-wave approximation \cite{saha}. While most potentials can
reproduce this value, the J\"ulich potential (J04) yields a stronger
binding while the old Nijmegen potential (NSC89) underestimates
the binding.

With the exception of the J04 and NSC89 potentials, all others agree up
to the saturation density. However, with increasing density, the
differences between these potentials grow, leading to different
bindings at rest. This will have consequences for the predictions of
the $\Lambda$ hyperon concentration in dense nuclear matter. In
particular, this will affect the maximum mass of neutron stars. It is
interesting to observe that even the Nijmegen potentials NSC97a-f are
different from each other at higher densities, since the only difference
between the bare potentials NSC97a-f is the $F/(F+D)$ ratio.

In the past, the potentials NSC89 \cite{schulze} and NSC97a,f
\cite{vidana1} have also been used as a basis for a single-particle
potential calculation in the $G$-matrix formalism. These $G$-matrix
calculations yield a more attractive $\Lambda$ potential. For example,
at saturation density a potential depth of $-29.8\;\text{MeV}$ is
found for the NSC89 potential, the NSC97a gives $-39.7\;\text{MeV}$,
and the NSC97f $-36.6\;\text{MeV}$. On the other hand, a comparison
with another $G$-matrix calculation \cite{rijken}, which uses a
different prescription for intermediate state spectra, yields similar
results to ours.

\begin{figure}[!ht]
  \centerline{\hbox{
  \includegraphics[width=1.3\linewidth]{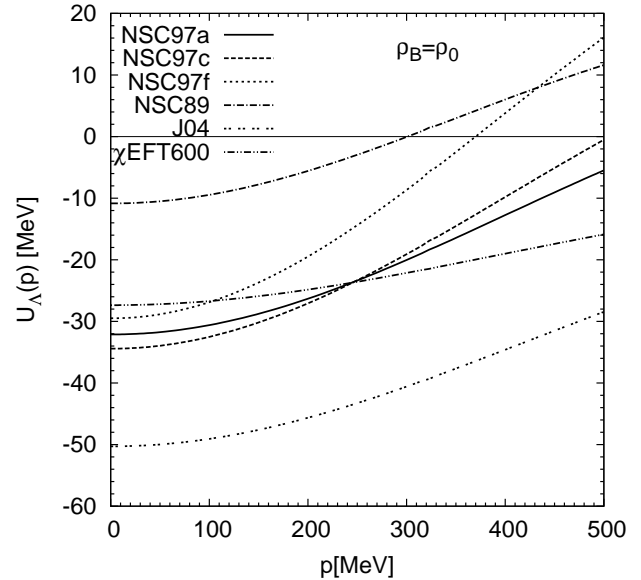}}}
\caption{Momentum dependence of $U_\Lambda (p)$ at saturation density
  in symmetric nuclear matter.}
  \label{fig:Lk}
\end{figure}

In \Fig{fig:Lk} the momentum dependence of the $\Lambda$ potential at
saturation density for various $YN$ potentials is shown. While all
potentials increase with increasing momentum, the slopes deviate of
each other. Similar differences in the momentum behavior of the
single-particle potentials are also seen in other works,
cf.~e.g.~\cite{schulze,vidana}.

\begin{table}[h!]
\begin{center}
\begin{tabular}{|c||c|c|c|c|c|c|c|c|}
\hline
       &$^1S_0$&$^3S_1$&$^1P_1$&$^3P_0$&$^3P_1$&$^3P_2$&$^3D_1$&$U_{\Lambda}$\\
\hline\hline
NSC97a & -4.86 &-27.79 & 1.70  &-0.10  & 2.10  & -2.03 & -0.09 &-32.12      \\
\hline
NSC97b & -6.69 &-27.40 & 1.86  & 0.05  & 2.53  & -1.87 & -0.09 &-32.72      \\
\hline
NSC97c & -9.06 &-27.54 & 1.96  & 0.36  & 2.84  & -1.72 & -0.09 &-34.42      \\
\hline
NSC97d &-12.14 &-26.05 & 2.22  & 0.64  & 3.54  & -1.33 & -0.08 &-34.46      \\
\hline
NSC97e &-13.92 &-24.43 & 2.43  & 0.75  & 4.09  & -1.03 & -0.07 &-33.50      \\
\hline
NSC97f &-15.37 &-20.85 & 2.85  & 0.68  & 5.09  & -0.47 & -0.05 &-29.49      \\
\hline
NSC89  &-15.73 &  4.52 & 2.00  & 0.52  & 2.55  & -3.46 & -0.07 &-10.84      \\
\hline
J04    & -9.55 &-35.18 &-0.15  &-0.70  & 0.58  & -3.17 & -1.31 &-50.28      \\
\hline
$\chi$EFT550&-11.11 &-15.46 & 1.50  &-1.69  & 3.17  & -0.07 & -3.14 &-27.14 \\
\hline
$\chi$EFT600&-12.29 &-11.39 & 1.50  &-1.73  & 3.17  & -0.07 & -6.14 &-27.37 \\
\hline
$\chi$EFT650&-11.99 & -6.70 & 1.50  &-1.77  & 3.17  & -0.07 & -9.90 &-26.27 \\
\hline
$\chi$EFT700&-11.91 & -1.77 & 1.50  &-1.81  & 3.17  & -0.08 & -13.84&-25.35 \\
\hline
\end{tabular}
\caption{Partial wave contributions to the $\Lambda$ potential
$U_{\Lambda}(p=0)$ at $\rho_B=\rho_0$ in symmetric nuclear matter.}
  \label{tab:Lpw}
\end{center}
\end{table}

In addition, \Eq{eq:spp} cannot only be used for the calculation of
$U_Y(p)$, but also to extract the individual partial-wave
contributions to the total potential. These contributions are obtained
by neglecting the summation over the $LSJ$ quantum numbers in
\Eq{eq:spp}, which we label in the following as $U_Y(^{2S+1}L_J)$. In
\Tab{tab:Lpw} and \Tab{tab:Smpw} the resulting partial-wave
contributions to $U_{\Lambda}$ and $U_{\Sigma^-}$, respectively, are
listed for several $YN$ interactions at vanishing momenta at
saturation density.

\begin{table}[h!]
\begin{center}
\begin{tabular}{|c||c|c|c|c|c|c|c|c|}
\hline
       &$^1S_0$&$^3S_1$&$^1P_1$&$^3P_0$&$^3P_1$&$^3P_2$&$^3D_1$&$U_{\Sigma^-}$\\
\hline\hline
NSC97a & 3.51  & -4.87 & -2.16 & 0.59  & 1.46  & -2.41 & -0.01 &-4.73       \\
\hline
NSC97b & 3.58  & -5.37 & -2.14 & 0.63  & 1.54  & -2.31 & -0.01 &-4.91       \\
\hline
NSC97c & 3.48  & -6.50 & -2.12 & 0.68  & 1.59  & -2.18 &  0.00 &-5.86       \\
\hline
NSC97d & 3.50  & -6.08 & -2.02 & 0.71  & 1.70  & -1.92 &  0.01 &-4.88       \\
\hline
NSC97e & 3.50  & -5.24 & -1.94 & 0.72  & 1.78  & -1.75 &  0.02 &-3.65       \\
\hline
NSC97f & 3.51  & -5.11 & -1.85 & 0.71  & 1.90  & -1.60 &  0.02 &-3.14       \\
\hline
NSC89  &-4.32  & 11.46 & -0.77 & 0.93  & 2.27  & -1.49 &  0.28 & 7.61       \\
\hline
J04    &-7.63  &  1.84 & -0.15 & 0.52  &-0.70  & -3.37 & -3.65 &-15.13       \\
\hline
$\chi$EFT550& 2.28  & 14.69 &  1.50 &-0.20  & 0.09  & -0.01 & -2.73 & 14.11 \\
\hline
$\chi$EFT600&-3.70  & 66.26 &  1.50 &-0.28  & 0.06  & -0.01 & -5.36 & 56.89  \\
\hline
$\chi$EFT650&-2.72  & 42.41 &  1.50 &-0.35  & 0.01  & -0.01 & -8.60 & 30.38  \\
\hline
$\chi$EFT700&-2.93  & 39.93 &  1.50 &-0.41  &-0.04  & -0.02 & -11.60& 24.68  \\
\hline
\end{tabular}
\caption{Partial-wave contributions to the $\Sigma^-$ single-particle potential
$U_{\Sigma^-}(p=0)$ at $\rho_B=\rho_0$.}
  \label{tab:Smpw}
\end{center}
\end{table}

In these tables the partial waves up to $L=2$ are shown and the last
column contains the sum up to $L=5$. As expected, the influence of the
$S$- and $D$-waves is the most dominant one. One can see that the
combination of the coupled $^3S_1$ and $^3D_1$ channels provides most
of the attraction in the majority of the $\Lambda$ potentials.

These tables also illustrate the different contributions to the
hyperon potentials originating from the central, spin-spin, spin-orbit
parts of the $YN$ interaction. These partial waves can than 
be use to determine the size of these contributions, which could 
than be compared to other results such as 
\cite{kaiserweise, kaiser, rijken, Pirner:1979mb}.
Furthermore, one can recognize that a
change in the $F/(F+D)$ ratio from the different bare NSC97a-f
potentials affects $U_\Lambda$ stronger than the $U_\Sigma$.

\begin{figure}[htbp]
  \centerline{\hbox{
      \includegraphics[width=\linewidth]{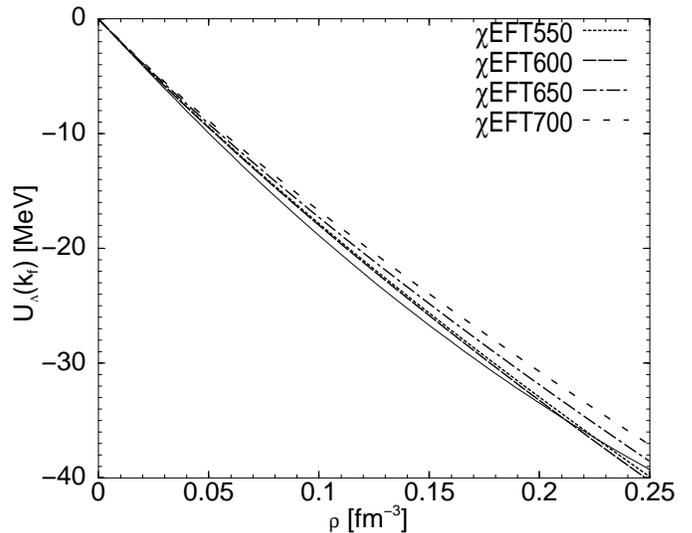}}}
  \caption{Density dependence of $U_{\Lambda}(p=0)$ for symmetric
    nuclear matter, $\rho=2k_F^3/3\pi^2$. Full lines are from
    \cite{kaiserweise} and dashed lines represent \eft for various
    regulator cutoffs.}
   \label{fig:kaiserL}
\end{figure}

Another interesting feature is that the \eft successfully reproduces
the potential depth at saturation density. For these densities, the
\eft agrees well with the Nijmegen NSC97a-f potentials.
\Fig{fig:kaiserL} shows a comparison of the $U_{\Lambda}(p=0)$ density
dependence, obtained with the \eft, with results from
\cite{kaiserweise}. Perfect agreement for the $U_{\Lambda}(p=0)$ is
evident and the independence of the \eft potential on the regulator
cutoff is also seen. This suggests that the two approaches in Refs.
\cite{polinder} and \cite{kaiserweise} to construct an \eft, are
closely related. Furthermore, \eft in leading order can already
produce a reasonable $\Lambda N$ potential.

\begin{figure}[!ht]
  \centerline{\hbox{\includegraphics[width=1.3\linewidth]{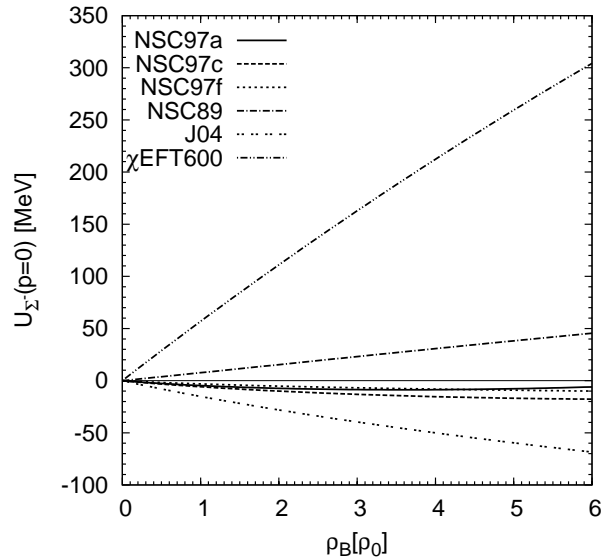}}}
  \caption{\label{fig:Smrho} Similar to \Fig{fig:Lrho} for the
    $U_{\Sigma^-} (p=0)$.}
\end{figure}

\Fig{fig:Smrho} shows the density dependence for several $\Sigma^-$
potentials at rest in symmetric nuclear matter similar to the previous
figure. The other members of the $\Sigma$ triplet, $\Sigma^+$ and
$\Sigma^0$, exhibit an almost identical behavior. A small differences
compared to the $\Sigma^-$ case is found to be due to a small
difference in their masses. Therefore, in the remaining section we
discuss only the $\Sigma^-$ potential. For $U_{\Sigma^-}$ no density
range is found where all or even most potentials agree. However, the
difference between the NSC97a-f potentials is not significant and is
the same over the entire range of densities shown. This further
confirms that the influence of the $F/(F+D)$ ratio on the $\Sigma N$
interaction is less important than on $\Lambda N$. Due to experimental
uncertainties in case of the $\Sigma$ potential, no generally accepted
empirical point can be used as a reference. On the one hand, recent
results \cite{saha}, based on the distorted-wave impulse
approximation, yield a repulsive potential of the order of
$100\; \text{MeV}$. On the other hand, the analysis of the same data
by \cite{Kohno:2006iq} in a semiclassical distorted-wave model and an
analysis by \cite{maekawa} within a distorted-wave impulse
approximation with a local optimal Fermi averaged $T$-matrix find a
less repulsive potential. In addition, there exists a bound state of
$^4_{\Sigma}He$ \cite{naege}, which definitely requires an attractive
potential. Thus, for this case no conclusive answer on the theoretical
as well as experimental side can be given.

Compared with a $G$-matrix calculation a stronger binding for the
$\Sigma$ single particle potential is found. In particular, using the
NSC89 interaction a binding of $-15.3\;\text{MeV}$ \cite{schulze} is
found while for the NSC97a potential $-29.7\;\text{MeV}$ and for the
NSC97f potential $-25.5\;\text{MeV}$ are reported \cite{vidana1}. In
order to understand the origin of such a significant difference the
individual partial-wave contributions to the potential by
\cite{schulze} are compared with each other. The $^1S_0$ channel
contributions are approximately the same while those for the $^3S_1$
channel are significantly different.

This difference in the $^3S_1$ channel is present for both $\Lambda$
and $\Sigma^-$ potentials and is the results of the difference in the
treatment of the $^3S_1$ $\Lambda N-\Sigma N$ channel. Since the
effective interactions are constructed from $\vlowk$ and the
$G$-matrix formalisms, respectively, the difference comes from the
treatment of the attractive part of the bare potentials above the
cutoff. This is similar to the point made in \Fig{fig:running3} where
a cutoff dependence is visible: by lowering the cutoff more and more
'attraction' is effectively added to the interaction. It is also
interesting to note, that the effective potentials constructed in the
$G$-matrix calculations, NSC89 and NSC97a,f depend on the underlying
bare potentials in a similar way as the potentials shown here. This is
another sign that the uncertainties are inherent in the underlying
potentials.

\begin{figure}[!ht]
  \centerline{\hbox{
  \includegraphics[width=1.3\linewidth]{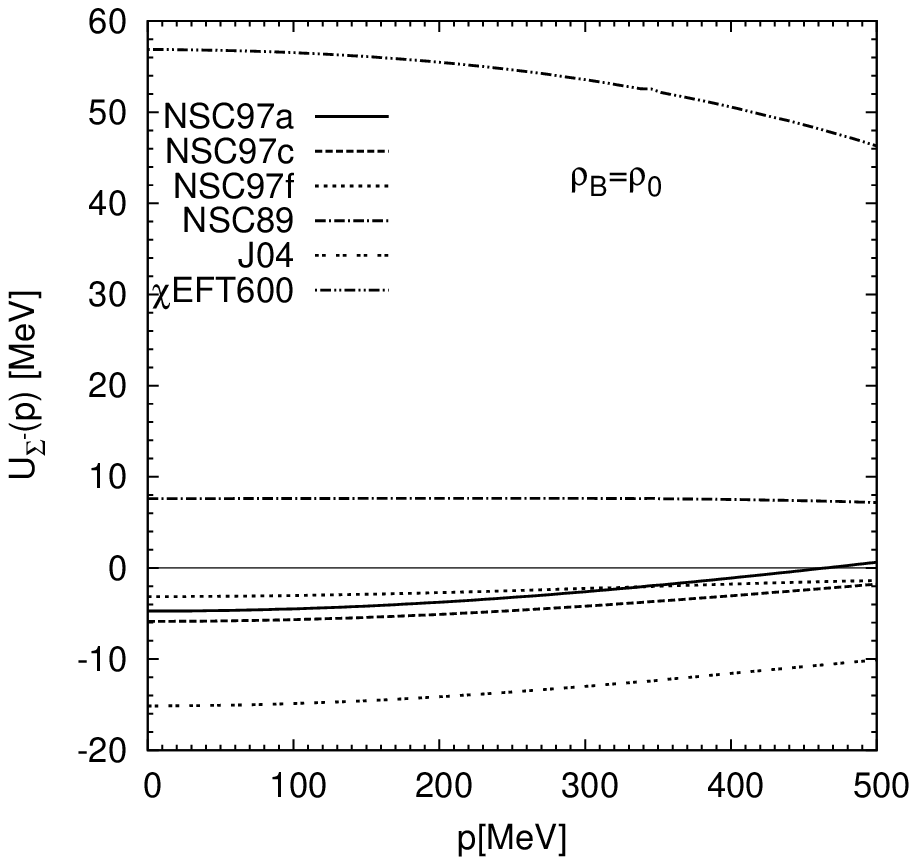}}}
  \caption{Similar to \Fig{fig:Lk} for $U_{\Sigma^-} (p)$.}
  \label{fig:Smk}
\end{figure}

In \Fig{fig:Smk} the momentum dependence of the $U_{\Sigma^-}$ at
saturation density for various $YN$ potentials is displayed and
illustrates how strongly the results depend on the parameterization of
the $YN$ interaction and reiterates that the $\Sigma N$ interaction is
very poorly constrained.

\begin{figure}[htbp]
  \centerline{\hbox{
      \includegraphics[width=\linewidth]{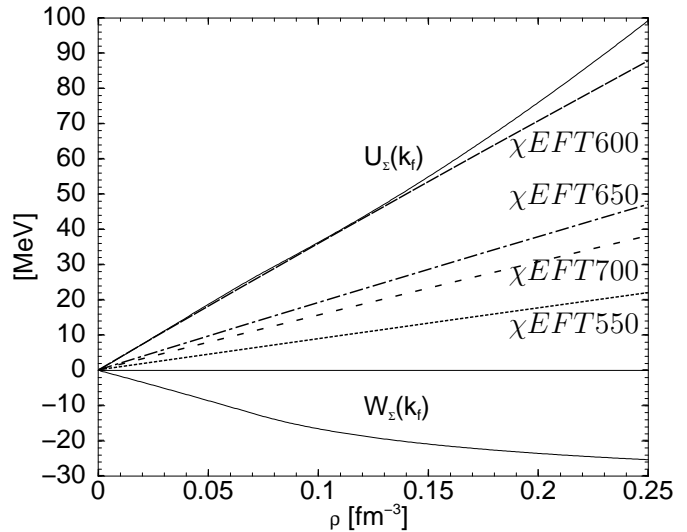}}}
  \caption{Density dependence of $U_{\Sigma}(p=0)$ for symmetric
    nuclear matter. The full line comes from \cite{kaiser} and the
    dashed lines represent \eft results for various values of the
    regulator cutoff.}
   \label{fig:kaiserS}
\end{figure}

\Fig{fig:kaiserS} shows the density dependence of the real part
$U_{\Sigma}(p=0)$ and the imaginary $W_{\Sigma}(p=0)$ of an optical
potential calculation of Ref.~\cite{kaiser} together with the results
obtained from \eft. The most interesting feature here is that all
potentials are positive and grow with increasing density in contrast
to other potentials. However, unlike in the case of the $\Lambda$
potential, $U_{\Sigma^-}$ depends on the regulator cutoff and only
\eft with a cutoff of 600 MeV agrees with the results of Ref.
\cite{kaiser} quantitatively. As already mentioned earlier, the
repulsive $\Sigma^-$ potential, which grows with density, has been
suggested by Saha et al.~\cite{saha} by means of an analysis of
($\pi^-,K^+$) inclusive spectra.

Recently, a calculation of the binding energy of the $\Lambda$ hyperon
in nuclear matter within a Dirac-Brueckner-Hartree-Fock framework was
performed using the most recent J\"ulich meson exchange $YN$ potential
\cite{Sammarruca:2008hy}. The reported values of the $\Lambda$
potential of $-51.27\;\text{MeV}$ ($-47.4\;\text{MeV}$) in
Brueckner-Hartree-Fock (Dirac-Brueckner-Hartree-Fock) framework agree
well with our prediction of $-50.28\;\text{MeV}$.

\section{Summary and conclusions}
\label{sec:summary}

In this paper an application of recently constructed $YN$ \vlowk\
potentials \cite{schaefer, wagner} is presented. The potentials were
constructed in a RG formalism and applied to nuclear matter. The
calculation of the single-particle potential of the $\Lambda$ and
$\Sigma$ hyperon was performed in the Hartree-Fock approximation.
Since the \vlowk\ is an low-momentum interaction, it can be directly
used in the considered range of momenta.

For comparison with other approaches, scattering lengths and the
single-particle potential of several bare potentials were calculated.
In contrast to the $NN$ case, it is not possible to obtain a precise
fit of the $YN$ potential to all partial waves, due to the incomplete
$YN$ scattering data base. This deficiency is clearly visible in the
results obtained.

A few interesting observations are in order, however. The \vlowk for
the $\Sigma^- N$ $^1S_0$ channel is cutoff independent in between
momenta of $250$ and $500$ MeV, but no such independence was found in
the $^3S_1$ channel due to a strongly attractive short-range part. The
results for the $\Lambda$ potential are more satisfying since most of
the potentials agree up to saturation density and reproduce the
empirical point. An interesting observation is that \eft in leading
order lies among these potentials. However, for the $\Sigma^-$
potential no such agreement exists. Unfortunately, the present
experimental situation for $\Sigma^-$ does not allow for more
stringent conclusions.

Finally, the results are also compared with other approaches,
\cite{schulze, vidana, rijken}. Most of the results of a $G$-matrix
calculations yield a stronger binding than we find. The difference
comes from the way in which the attraction, especially in the
$\Sigma N\rightarrow \Lambda N$ channels, is treated in the RG
approach. In comparison with \eft overall agreement is seen
\cite{kaiserweise, kaiser}.

However, the most interesting fact is that differences
in approaches, \vlowk, G-matrix and \eft, are much less than the 
differences between the various potentials. Thus the differences
we see here are not the result of the approaches used to construct
the effective interaction, but largely reflect uncertainties in
the underlying bare potentials.

\section*{Acknowledgments}
\label{sec:Acknowledgements}

This work has been partially supported by the Helmholtz Gemeinschaft,
program Grant No. VH-VI-041 and by the Helmholtz Research School for
Quark Matter Studies. We would like to thank Mathias Wagner and Isaac
Vida\~{n}a for useful discussions.


\end{document}